\documentclass[conference]{IEEEtran}
\IEEEoverridecommandlockouts
\usepackage{cite}
\usepackage{lipsum}  
\usepackage{algorithmic}
\usepackage{graphicx}
\usepackage{textcomp}
\usepackage{xcolor}
\usepackage[left=1.62cm,right=1.62cm,top=1.9cm]{geometry}
\usepackage{amsmath,amssymb,amsfonts,amsthm}
\usepackage{bbm}
\usepackage{mathtools}
\usepackage{comment}

\DeclarePairedDelimiter\abs{\lvert}{\rvert}

\usepackage{xcolor}
\usepackage{bm}
\usepackage{array}
\newcounter{thm}

\newtheorem{definition}[thm]{Definition}

\newcommand*{\rom}[1]{\uppercase\expandafter{\romannumeral #1\relax}}

\newcommand{\ket}[1]{{|#1\rangle}}   
\newcommand{\bra}[1]{{\langle#1|}}

\newcommand{\tr}{{\mathrm{tr}}}

\allowdisplaybreaks

\makeatletter

\counterwithout{equation}{section}
\def\BibTeX{{\rm B\kern-.05em{\sc i\kern-.025em b}\kern-.08em
    T\kern-.1667em\lower.7ex\hbox{E}\kern-.125emX}}
\begin{document}

\title{Trading Datarate for Latency in Quantum Communication
}

\author{
   \IEEEauthorblockN{Zuhra Amiri\IEEEauthorrefmark{1}, Florian Seitz\IEEEauthorrefmark{1}, Janis N\"otzel\IEEEauthorrefmark{1}}
   \IEEEauthorblockA{
       \IEEEauthorrefmark{1}Emmy-Noether Group Theoretical Quantum Systems Design Lehrstuhl f\"ur Theoretische Informationstechnik,\\ Technische Universit\"at M\"unchen\\ 
  \{zuhra.amiri, flo.seitz,  janis.noetzel\}@tum.de 
    }}

\maketitle

\begin{abstract}
Low latency and high data rate performance are essential in wireless communication systems. This paper explores trade-offs between latency and data rates for optical wireless communication. We introduce a latency-optimized model utilizing compound codes as one corner case and a data rate-optimized model employing channel estimation via pilot signals and feedback before data transmission. Trade-offs between the two extremes are displayed. Most importantly, we detail operating points that can only be reached when the receiver side of the link employs optimal quantum measurement strategies.
Furthermore, we propose an IoT application in a robot factory as an example scenario. Our findings reveal a trade-off between latency and data rate driven by two basic algorithms: compound codes reduce latency at the cost of data rates, while channel estimation enhances data rates at the cost of latency.
\end{abstract}

\begin{IEEEkeywords}
optical computation, optical wireless communication, low latency
\end{IEEEkeywords}

\section{Introduction}

In communication systems, the primary objectives are lowering latency and increasing data rates — particularly in applications related to the Internet of Things (IoT) \cite{aleksic2019survey}. Recent studies have explored Optical Wireless Communication (OWC) as a promising alternative to traditional Radio Frequency Communication (RFC) for achieving these goals \cite{aleksic2019survey, khalighi_survey_2014}. OWC offers significant advantages in terms of faster data rates and lower latency, making it well-suited for high-demand IoT applications \cite{khalighi_survey_2014}. For example, in telesurgery, where real-time data transmission is critical for precision and safety, OWC can provide the low-latency communication required for remote surgical operations \cite{chen2018ultra}. In intelligent transportation systems, OWC can enhance the speed and reliability of data exchange between vehicles and infrastructure, improving traffic management and safety \cite{chen2018ultra}. Additionally, in industrial automation, OWC can support the high-speed data requirements of advanced manufacturing processes and machinery \cite{chen2018ultra}.
However, despite the benefits of OWC, it faces challenges, such as the need for precise alignment between optical transmitters and receivers. This alignment issue, often called the line-of-sight (LOS) path, can be a significant obstacle. To address these challenges, various solutions are proposed, including channel estimation techniques to track and align the transmitter and receiver for the LOS path. Alternatively, a non-line-of-sight (NLOS) path can be employed by spreading the beam using a lens \cite{Cui2010}. While this approach can mitigate alignment issues, it has the drawback of reducing the photon count due to the beam's spread, which may impact the communication quality in scenarios like intelligent transportation where high data rates are crucial.
Moreover, emerging technologies such as Surface Plasmon Amplification by Stimulated Emission of Radiation (SPASERs) offer additional promise. SPASERs, functioning similarly to lasers but on a much smaller scale, present unique advantages such as low power consumption and potential suitability for energy-efficient systems. These characteristics are particularly relevant in contexts like industrial automation, where energy efficiency is a priority. However, the low output power of SPASERs introduces challenges that can be addressed through quantum communication techniques, known for their advantages in low photon number scenarios \cite{noetzelRosati}. Quantum receivers, for example, can improve the performance of communication systems where the photon count is low, thus complementing the strengths of SPASERs and addressing their limitations.
In summary, while OWC and SPASERs represent significant advancements in communication technology, their integration with quantum communication techniques offers a pathway to overcoming current limitations and enhancing performance across various IoT applications, including telesurgery, intelligent transportation, and industrial automation.

\begin{figure}[h!]
    \centering
    \includegraphics[scale = 0.35]{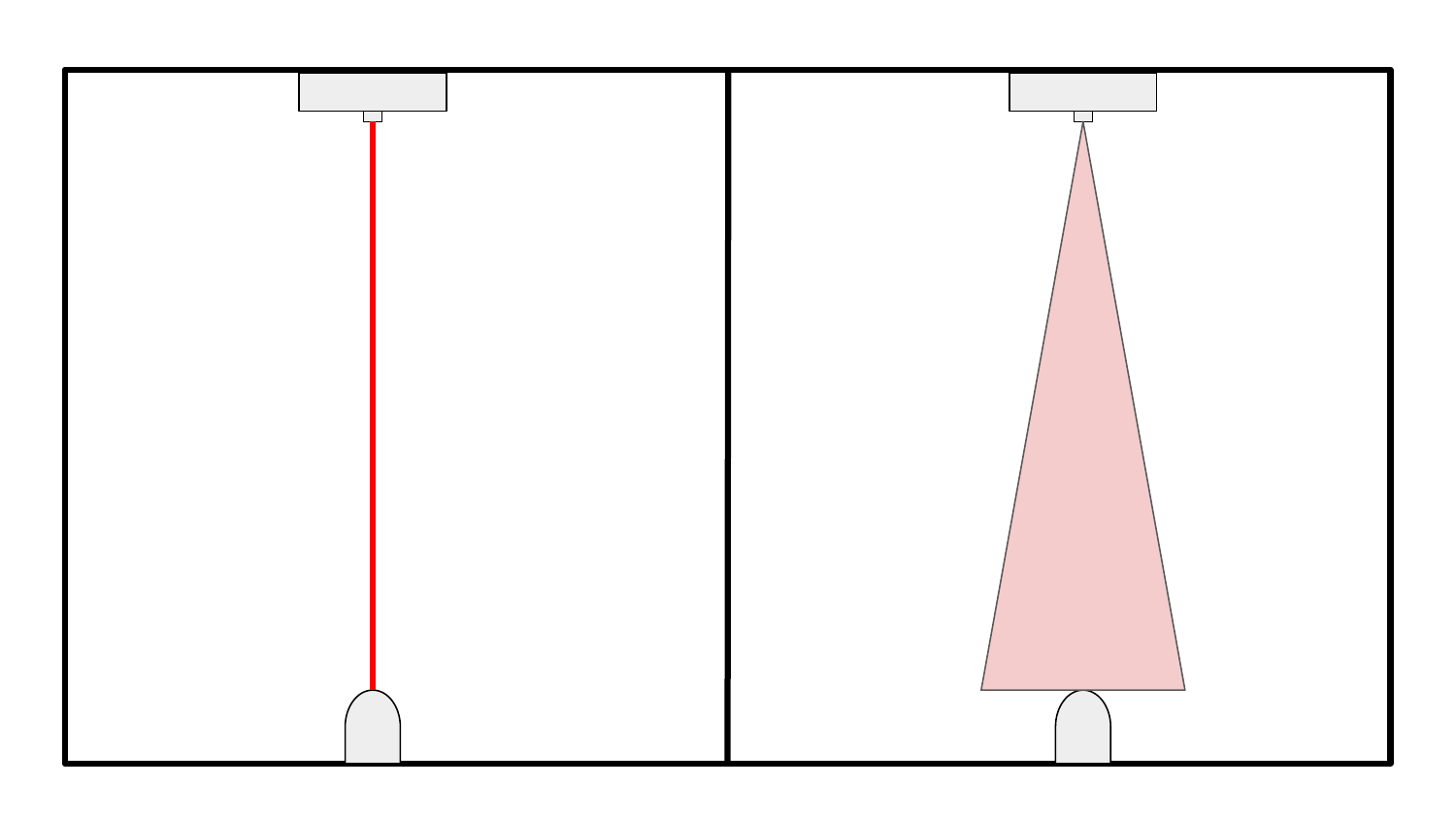}
    \caption{The left-hand side shows a LOS path, where the transmitter and the receiver are perfectly aligned. The right-hand side shows an NLOS path, where the transmitter and receiver do not have to be perfectly aligned due to the broadened beam.}
    \label{fig:losnlos}
\end{figure}

Other significant challenges include atmospheric effects — like rain, fog, and turbulence — and lights' inability to penetrate obstacles — like walls, people, and other objects \cite{hamza_optical_2020, Yuan2016}. 
Because of these challenges, the channel parameters can vary vastly. There are two ways to combat this problem. Firstly, one can estimate the channel by sending pilot signals, which come at the cost of latency; secondly, one can use compound codes. While the compound code may reduce the data rate, they are a class of coding schemes designed to perform well under uncertainty about the communication channel. Unlike traditional codes that require accurate channel state information (CSI), compound codes are robust to variations in the channel. This makes them ideal in scenarios where CSI is unavailable or too costly. Other codes may rely heavily on precise channel knowledge, leading to performance degradation when the channel is poorly understood or changes frequently. 
In this paper, we aim to explore the capacity of a classical-quantum compound channel. Firstly, we will introduce system models for the pure loss bosonic channel, with one model optimized for latency and the other for data rate. We then detail one method for the estimation of the channel loss parameter. Subsequently, we outline potential applications of our proposed models. Finally, we present our findings, provide a comprehensive discussion, and conclude with avenues for future research.

\section{Notation}

We model our systems on the Fock space denoted as $\mathcal F=\mathrm{span}(\{|r\rangle\}_{r\in\mathbb N})$. The signal class of the transmitter is that of coherent states $|\alpha\rangle=e^{-|\alpha|^2/2}\sum_{r=0}^\infty\alpha^r/\sqrt{r!}$, $\alpha\in\mathbb C$, where $|\alpha|^2$ is proportional to the signal energy. For a given transmissivity $\tau\in[0,1]$ of the channel, the receiver receives $|\sqrt{\tau}\alpha\rangle$ if the sender emitted the signal $|\alpha\rangle$. The trace of an operator $A$ is needed to model the detection process and is denoted as $\tr(A)$. The identity operator is $\mathbbm1$. The logarithm $\log$ is taken with respect to base two. Classical-quantum channels will be denoted as $\mathcal N$. The set of all such channels, with input set $\mathcal X$ and output space $\mathcal H$, is $C(\mathcal X,\mathcal H)$.

In this work, we combine two elementary tasks. The first is coding, which in our notation accepts an input parameter $\epsilon>0$ and an interval $I$ and then produces a sequence $(b_n)_{n\in\mathbb N}$ where $b_n$ is the number of bits transmitted in $n$ channel uses with error no more than $\epsilon$, given that the channel parameter $\tau$ satisfies $\tau\in I$.

The second is estimation. An estimator in our setting is a measurement setup which, given parameter $\epsilon>0$, produces a sequence of intervals $(I_{n,\hat\tau})_{n\in\mathbb N,\hat\tau\in[a,b]}$ such that for all possible transmissivity values $\tau\in[a,b]$ it holds that if the estimator is run on $n$  outputs of the channel characterized by $\tau$, producing a probability distribution $p(\cdot|\tau)$, then $p(I_{n,\hat\tau}|\tau)\geq1-\epsilon$.

We combine both tasks to create codes with one round of feedback where decoding operations take the form $(D_{\tau}^{n_1} \otimes F_{m,\tau}^{n_2})_{m=1,\tau\in T}^{M_\tau}$ where $T = [a,b]\subset(0,1]$ is an interval of channel states, $1,\ldots,M_\tau$ are messages and $n_1,n_2\in\mathbb N$ are the length of the the channel estimation block and the data transmission block, respectively. The requirement on the codes is that the construction should not depend on $\tau\in T$, and the probability of successful transmission should satisfy
\begin{align}
    &\min_{\tau\in[a,b]}\int_a^b p_{D,\hat\tau,\tau}\cdot p_{E,\hat\tau,\tau}  d \hat\tau \geq1-2\epsilon,
\end{align}
where 
\begin{align}
    p_{D,\hat\tau,\tau}&:=\tr(D_{\hat\tau}^{n_1(\hat\tau)}N_\tau(\Hat{\sigma}))\\
    p_{E,\hat\tau,\tau}&:=\frac{1}{M_{\hat\tau}}\sum_{m=1}^{M_{\hat\tau}}\tr(F_{m,\hat\tau}^{n_2(\hat\tau)}\mathcal N_\tau(\rho_{\hat\tau,m}),
\end{align}
$\Hat{\sigma}$ is the pilot state used for channel estimation, and $\rho_{\hat\tau,m}$ is the state encoding the message $m$ when the channel transmissivity is $\Hat{\tau}$.
For every $\tau\in[a,b]$, the latency of the code is measured as 
\begin{align}\label{eqn:latency-formula}
    n(\tau)=\int_a^b\tr(D_{\hat\tau}^{n_1(\hat\tau)}\Hat{\sigma})(n_1(\hat\tau)+n_2(\hat\tau)) d\hat\tau,
\end{align}
and the data rate as
\begin{align}\label{eqn:data-rate-formula}
d(\tau)=\int_a^b\tr(D_{\hat\tau}^{n_1(\hat\tau)}\Hat{\sigma})\tfrac{1}{n_1(\hat\tau)+n_2(\hat\tau)}\log M_{\hat\tau}d\hat\tau.
\end{align}
The latency-data-rate region is the set of all tuples $(n(\tau),d(\tau))$.

\section{Definitions}
In this section, we will show essential definitions. The quantum second-order coding rate is taken from \cite{Wilde_2015}, and the classical counterpart from \cite{Polyanskiy2010}.

\begin{definition}[Quantum Second-Order Coding Rate]
The second-order coding rate gives the number of bits one can send over $n$ uses of a pure-loss bosonic channel with failure probability no larger than $\epsilon \in (0,1)$. The lower bound for sufficiently large yet finite $n$ is given by
 \begin{align}\label{eqn:second-order-coding-rate}
    \log M^*(&\mathcal{N}_{\eta}^{\otimes n}, N_S, \epsilon) \geq \nonumber\\
    &n g( \eta N_S) + \sqrt{nv(\eta N_S)}\Phi^{-1}(\epsilon)+O(\log n),
\end{align}

where $M^*(\mathcal{N}_{\eta}^{\otimes n}, N_S,\epsilon)$ is the maximum number of messages that can be transmitted over a bosonic channel $\mathcal{N}$ with photon number constraint $N_S \in [0, \infty)$ and $\eta \in (0,1]$ is the channel loss parameter. 

Furthermore, $g(\eta N_S)$ is the entropy of thermal state with mean photon number $\eta N_S$, given by 
\begin{align}
    g(x)=(x+1)\log(x+1)-x\log x, 
\end{align}
$\Phi^{-1}(\epsilon)$ is the inverse of the cumulative normal distribution function, and $v(\eta N_S)$ is the entropy variance, given by
\begin{align}
    v(x) = x(x+1)[\log(x+1)-\log x]^2.
\end{align}
    
\end{definition}

\begin{definition}[Classical Second-Order Coding Rate]
The second-order coding rate gives the number of bits one can send over $n$ uses of an AWGN channel with SNR $P$ and with failure probability $\epsilon \in (0,1)$. The approximation for block length $n$ is given by
     \begin{align}
         \log M^*(n,\epsilon, P) = n\cdot C - \sqrt{nV}Q^{-1}(\epsilon)+O(\log n),
     \end{align}
     where $C$ is the capacity $C = \frac{1}{2}\log(1+P)$, $V$ is the channel dispersion $V = \frac{P}{2}\frac{P+2}{(P+1)^2}\log^2e$ and $Q$ is the complementary Gaussian cumulative distribution function. 
\end{definition}

\begin{definition}[POVM]
  A positive operator-valued measure (POVM) is a set $\{ \Lambda_j\}_j$ of operators that satisfy non-negativity and completeness:
  \begin{align}
      \forall j: \Lambda_j \geq 0 \quad \quad \sum_j \Lambda_j = \mathbbm{1}
  \end{align}
  The probability for obtaining outcome $j$ is $\bra{\psi}\Lambda_j\ket{\psi}$, if the state is some pure state $\ket{\psi}$. The probability for obtaining outcome $j$ is $\tr(\Lambda_j\rho)$ if the state is in a mixed state described by some density operator $\rho$. \cite{wilde2013quantum}

\end{definition}

\begin{definition}[Photon Number in a Laser Pulse]
    The photon number in a laser pulse is given by
\begin{align}
    N_S = \frac{P}{E},
\end{align}
with $P$ being the power of the laser and $E$ being the energy of one photon at wavelength $\lambda$ given by
\begin{align}
    E = \frac{hc}{\lambda}
\end{align}
\end{definition}

\section{System Model}

This section will introduce two system models we have investigated — the latency optimal system model and the data-rate optimal system model. Furthermore, we will explain the channel estimation and give examples of applications.

Let $\mathcal N = \{ \mathcal N_\tau \}_{\tau \in \mathbf S}$ be a compound channel with state set $\mathbf S\subset[0,1]$ where
\begin{align}
    \mathcal N_\tau (\alpha) = \ket{\sqrt{\tau} \alpha}.
\end{align}
The values $\tau$ are called the transmissivity of the channel. A typical choice for $\mathbf S$ will be $\mathbf S=(a,b]$, but it is not required to be an intervall.

The transmissivity is given by the channel losses in free space optical communication — path loss, atmospheric absorption, scattering, diffusion turbulence, weather, and pointing loss. 

\subsection{Latency Optimal System Model}

This section presents the optimal latency system and analyzes its decoding latency and data rate. The system parameters include transmission power $P$, block length $n$, symbol rate $B$, and error rate $\epsilon$. 

The transmitter operates based on the worst-case assumption of distance $d$, which represents the maximum possible distance between the transmitter and receiver rather than the actual distance $d'$. This approach eliminates the need for pilot signals and allows for immediate data transmission. The transmitter begins sending data with an omnidirectional signal from an unknown location.
    
A lens can be used to achieve an omnidirectional beam \cite[p. 33]{hranilovic2006wireless}, \cite{elgala}. This beam divergence results in a lower photon density, as the photons are distributed over a wider area. 
    
Subsequently, the receiver utilizes a compound code as described in \cite{seitzNoetzel,cnr21,bbjn2012}. For the sake of simplicity, in this preliminary analysis, we use the reasonable assumption that the second-order coding rate of the compound channel is given by the worst-case second-order coding rate of the channels making up the compound channel.

The system is used without feedback, meaning data transmission will silently fail if $d'>d$. 

\subsection{Data Rate Optimal System Model}
In this section, we introduce the data rate latency system and analyze its decoding latency and data rate. The system parameters include transmission power $P$, block length $n$, symbol rate $B$, and error rate $\epsilon$.
The transmitter assumes a worst-case scenario for the distance $d$ to the receiver. It sends an omnidirectional pilot signal with $n_p=\alpha_1, ..., \alpha_n$ pilot symbols from an unknown location, where $\alpha_i$ is $+\alpha$ or $-\alpha$ with probability $0.5$. The receiver then estimates the distance $d$ by generating an estimate $\hat d$ of the transmitted signal from the pilot signals. 

Subsequently, the receiver sends this information $(\hat d)$ back to the transmitter. We assume the receiver has unlimited transmission power, so the feedback is error-free.
For this study, we simplify the analysis by assuming that the worst-case channel dictates the second-order coding rate. Using Eq. \eqref{eqn:second-order-coding-rate}, we can calculate the number of bits that can be transmitted. After the transmitter receives the feedback, it can send data with a narrow, more concentrated beam to the receiver. Depending on the distance, the transmitter can adjust its baud rate to ensure a reliable data transmission with an error rate $\epsilon$.

\subsection{Channel Estimation of a Pure Loss Bosonic Channel Using Classical Methods}
\label{subsec:estimation}
This section will explore in detail a method to estimate a pure loss bosonic channel: namely, this will be homodyne detection together with classical post-processing.

Consider the task of determining the transmissivity $\tau$ of a pure loss bosonic channel through measurements of the pilot symbols $\ket{\sqrt{\tau E}}$ using only classical methods — in this case homodyne detection and classical post-processing. The outcome probability distribution of a homodyne measurement corresponding to the position quadrature operator on a coherent state $\ket{\alpha}$ with $\alpha \in \mathbbm{R}$ is
\begin{align} \label{eq:homodyne_pdf}
    p(u) = \frac{1}{\sqrt{\pi}} e^{-(u-\sqrt{2} \alpha)^2},
\end{align}
which is a real normal distribution $N(\mu,\sigma^2)$ with mean $\mu=\sqrt{2}\alpha$ and variance $\sigma = \sqrt{1/2}$. 
Therefore, the channel estimation task is reduced to a classical hypothesis testing problem for a Gaussian with known variance.

When sampling a normal distribution, the probability that the sample mean after $n$ samples $\Hat{\mu}_n$ deviates less than $\delta$ from the population mean is
\begin{align}
    P\left( \abs{\Hat{\mu}_n - \mu} < \delta \right) = 2 \Phi\left( \frac{\delta \sqrt{n}}{\sigma} \right) - 1,
\end{align}
where $\Phi$ is the cumulative distribution function (CDF) of the standard normal distribution. If we require this success probability to be larger than $1-\epsilon$ for some $\epsilon>0$ as well as a number of channel uses $n_1$, we arrive at 
\begin{align}
    \delta = n_1^{-1}\sqrt{1/2}\phi^{-1}(1-\epsilon/2)
\end{align}
as the uncertainty range of our prediction of the value $\sqrt{2\tau E}$ is now $[\sqrt{2\tau E}-\delta,\sqrt{2\tau E}+\delta]$, which translates to an uncertainty of $[\sqrt{\tau}-\delta/\sqrt{2 E},\sqrt{\tau}+\delta/\sqrt{2E}]$ for the estimate of $\sqrt{\tau}$ and finally setting $\delta'=\delta/\sqrt{2 E}$ to an uncertainty of 
\begin{align}
    [(\sqrt{\tau}-\delta')^2,(\sqrt{\tau}+\delta')^2] 
        &\subset [\tau-3\delta',\tau+3\delta']
\end{align}
as long as $\delta'<1$. For the derivation of our numerical results, we will later use $E>10^4/2$, so that $\sqrt{2 E}>10^2$ and thus $\delta'<\delta\cdot10^2$. Further we will use $\epsilon\geq10^{-5}/2$, and since $\sqrt{1/2}\phi^{-1}(1-10^{-5})<3$ our approximation of the uncertainty in our estimate of $\tau$ is valid for all $n_1\geq3$. 
\subsection{Construction of Estimator} \label{sec:construct_estimator}
    Let $\left\{M_u\right\}_{u \in \mathbbm R}$ be the POVM corresponding to the Homodyne measurement such that it produces the outcome distribution \eqref{eq:homodyne_pdf}. We define the set $S_{\hat\tau}:=\{\mathbbm R^{n_1} \ni q^{n_1}:\sum_iq_i=n_1\cdot\sqrt{2E\hat\tau}\}$ and
    \begin{align}
        D_{\hat\tau}^{n_1}:=\int_{q^{n_1}\in S_{\hat\tau}} M_{q_1} \otimes \dots \otimes M_{q_{n_1}} dq^{n_1}
    \end{align}
    For $\delta:=n_1^{-1}\Phi^{-1}(1-\epsilon/2)$ it then holds that
    \begin{align}
        \min_{\tau\in[a,b]}\int_{\tau-\delta}^{\tau+\delta}\tr(D_{\hat\tau}^{n_1}|\sqrt{\tau E}\rangle\langle\sqrt{\tau E}|^{\otimes n_1})d\hat\tau\geq1-\epsilon,
    \end{align}
    so that a prescribed maximum estimation error of $\epsilon$ and a number $n_1$ of channel uses together with a measured value $\hat\tau$ directly translate into an interval $I=[\hat\tau-6\delta',\hat\tau]$ which can be set as the parameter region for the use of a compound channel code during the next $n_2$ transmissions.

\subsection{Concatenated Code}
    For all channel parameters $\tau$, it then holds
    \begin{align}
        \int_a^bp_{D,\hat\tau,\tau}\cdot p_{E,\hat\tau,\tau}d\hat\tau 
            &\geq \int_{\tau-3\delta'}^{\tau+3\delta'}p_{D,\hat\tau,\tau}\cdot p_{E,\hat\tau,\tau}d\hat\tau\\
            &\geq \int_{\tau-3\delta'}^{\tau+3\delta'}(1-\epsilon)\cdot p_{E,\hat\tau,\tau}d\hat\tau\\
            &\geq (1-\epsilon)\cdot (1-\epsilon)\\
            &\geq1-2\epsilon,
    \end{align}
    where for the channel estimation strategy from Subsection \ref{subsec:estimation} and for the data transmission, a hypothetical compound code tailored to work for all channels in the interval $[\hat\tau-6\delta',\hat\tau]$ operating at data rate $\log M^*(\mathcal{N}_{\hat\tau-6\delta'}^{\otimes n_2}, E, \epsilon)$ where $n_2$ is the number of channel uses for data transmission. Upon fixing any two values $n_1,n_2\in\mathbb N$ with $n_1\geq3$, the data rate of the code is then given by
    \begin{align}
        d(\tau)=\int_a^bp_{E,\hat\tau,\tau}\tfrac{1}{n_1+n_2}\log M^*(\mathcal{N}_{\hat\tau-6\delta'}^{\otimes n_2}, E, \epsilon)d\hat\tau
    \end{align}
    and the latency evaluates to $n_1+n_2$, where we first used the property of the compound code, then the Asymptotic Equipartition Property (AEP) for the channel estimation, and finally $\epsilon^2>0$.
    
\subsection{Applications}
Potential applications for these system models can be in IoT, e.g., in a highly mobile robot factory. In this setting, tracking the transmitter is challenging; therefore, sending an omnidirectional beam is easier. We will investigate the trade-off between latency and data rate. 

Since we are investigating wireless optical communication systems in an indoor factory setting, this allows us to eliminate background noise, for example, by using light filters at the windows. Therefore, in our model, we only consider the free space path loss, which is given by 
\begin{align}
 \tau = \left(\frac{\lambda}{4 \pi d}\right)^2,  
\end{align}
where $d$ is the distance between transmitter and receiver \cite[p. 1321]{whitaker2018electronics}.

Moreover, saving costs and being as energy-efficient as possible in an IoT setting are essential. Therefore, a low-power laser should be utilized, e.g., $1mW$ to $100mW$.

\section{Results}
This section presents the results of our simulation. We set $[a,b]=[0.001,1]$, $E=10^4/2$ and $\epsilon=0.5\cdot10^{-5}$. We then considered two feedback-based systems, which use the channel estimation method based on homodyne detection as described in \ref{sec:construct_estimator}. We compare the achievable data rate–latency regions to the compound data rate, equivalent to the quantum scheme with $n_1$ fixed to zero and therefore $\hat\tau = a$.

\begin{figure}
        \centering
        \includegraphics[scale=0.41]{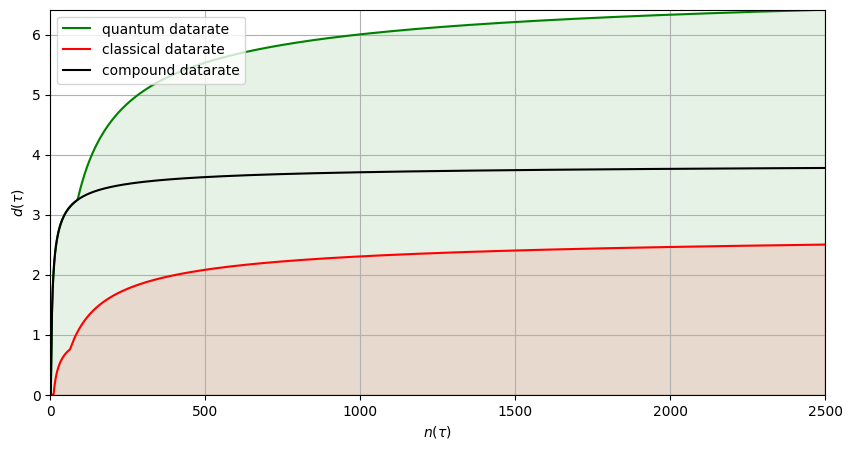}
        \caption{Datarate $d(\tau)$ vs latency $n(\tau)$ of a classical and a quantum sender-receiver pair, and the quantum compound case without any feedback for comparison. The energy is $E=10^4/2$, the transmissivity is $\tau = 0.01$, and the lower end of the parameter interval is $a = 0.001$.}
        \label{fig:plot1}
    \end{figure}

Figure \ref{fig:plot1} shows the simulation results. The transmissivity parameter used to calculate latency and data rate according to \eqref{eqn:latency-formula} and \eqref{eqn:data-rate-formula} was $\tau = 0.01$. 

We can see that the compound capacity is optimal for short block lengths. The transition point where channel estimation begins to provide an advantage is marked by a distinctive inflection point in the curves. 

In Figure \ref{fig:plot1}, we can observe that the classical data rate and the compound code reach their maximum after approximately $1000$ channel uses, while the compound data rate is higher than the classical rate. Additionally, it's evident that the quantum data rate continues to increase steadily even after $2500$ channel uses.

\section{Discussion}
In this section, we will discuss our findings.

The findings from the simulations show that while channel estimation and feedback mechanisms are instrumental strategies in enhancing the data rate of wireless optical communication systems, compound codes can significantly reduce latency. 

The initial channel estimation phase also consumes extra resources, such as power for transmitting pilot signals and feedback and computational power for averaging received signal strengths and estimating. Notably, the feedback mechanism introduces latency, causing delays in data transmission and reducing efficiency in the initial setup. If low latency is crucial for a communication system, the feedback mechanism may not be suitable. 

In contrast, compound codes are robust against multiple channel parameters and offer significant latency reduction at the price of data rate.

The results illustrate that the latency can be further reduced with quantum measurement techniques compared to classical methods.
Moreover, it is speculated that quantum measurements can further reduce the latency for systems with feedback mechanisms and provide more efficient estimations than classical methods. Using quantum techniques can be advantageous in applications with low photon numbers when low latency and high data rates are critical. 

While our paper introduces an application, more detailed use cases would help us understand the proposed models' practical implications. Use cases for healthcare systems or autonomous vehicles could provide further insight into the applicability of OWC system models. 

Moreover, in this study, we assumed the worst-case second-order coding rate to be reasonable. For future work, a proof of second-order coding rate for compound codes is imperative for the understanding of trade-offs between latency and data rates. 

Our channel estimation method relies on homodyne detection, which outperforms heterodyne detection in case the channel has a known phase relation. Our study focused on low-noise environments, assuming an indoor factory setting to eliminate background noise. This critical assumption enables our systems to operate under optimal conditions. Further work must, therefore, investigate thermal or phase noise, highlighting the importance of homodyne detection. Finally, the physical realization of the optimal quantum receiver is a major obstacle preventing access to a large part of the latency–data rate plane.

\section{Conclusion}

This paper investigates the trade-offs between latency and data rates in OWC systems, proposing two models: a latency-optimized model using compound codes and a data rate-optimized model employing channel estimation and feedback mechanisms once with quantum and once with classical methods. The findings indicate that compound codes are advantageous in scenarios requiring low latency, as they are robust against varying channel parameters but result in lower data rates. Conversely, channel estimation techniques enhance data rates at the cost of increased latency due to the feedback mechanism. A better estimation of the transmitter's position is crucial for link adaptation, e.g., estimating the distance and the angle of arrival. 
The study also speculates the potential benefits of quantum measurement techniques that improve channel estimation accuracy and maintain low latency while achieving high data rates. The simulations demonstrate these trade-offs, with quantum methods promising future research.

\section{Acknowledgments}
The authors acknowledge the financial support by the Federal Ministry of Education and Research of Germany in the programme of “Souver\"an. Digital. Vernetzt.”. Joint project 6G-life, project identification number: 16KISK002, and further under grant numbers 16KISQ093, 16KISR026, 16KISQ077. Further funding was received from the DFG in the Emmy-Noether program under grant number NO 1129/2-1, and by the Bavarian state government via the 6GQT and NeQuS projects. Support was provided by the Munich Center for Quantum Science and Technology (MCQST).
The research is part of the Munich Quantum Valley, supported by the Bavarian state government with funds from the Hightech Agenda Bayern Plus.

\bibliographystyle{IEEEtran}
\bibliography{references}

\end{document}